\documentclass[11pt]{revtex4}
\usepackage{color}
\usepackage{amsmath,amssymb}
\usepackage[dvipdfm]{graphicx}
\usepackage{cancel}

\allowdisplaybreaks

\begin{document}
\title{Interaction solutions for supersymmetric mKdV-B equation}

\author{Bo Ren$^{1}$\footnote{Corresponding author, e-mail: renbo0012@163.com.} }

\affiliation{$^1$Institute of Nonlinear Science, Shaoxing University, Shaoxing 312000, China }

\date{\today $\vphantom{\bigg|_{\bigg|}^|}$}

\begin{abstract}
The ${\cal N} =1$ supersymmetric mKdV-B system is transformed to a system of coupled bosonic equations by using the bosonization approach. The bosonized supersymmetric mKdV-B (BSmKdV-B) equation includes the usual mKdV equation and a linear partial differential equation. The bosonization approach can thus effectively avoid difficulties caused by anticommutative fermionic fields of the supersymmetric systems. The consistent tanh expansion (CTE) method is applied to the BSmKdV-B equation. A nonauto-B\"{a}cklund (BT) theorem is obtained by using CTE method.
The interaction solutions among solitons and other complicated waves including Painlev\'{e} II waves and periodic cnoidal waves are given through a nonauto-BT theorem. The features of the soliton-cnoidal interaction solutions are investigated both in analytical and graphical ways by combining the mapping and deformation method.

\end{abstract}

\maketitle



\section{Introduction}
The study of supersymmetric integrable systems has developed great importance in recent years \cite{Kupershmidt, Martin, Math, Kupershmid}.
The supersymmetrization of a number of integrable equations in which
the bosonic equation is independent of the fermionic variable and the system
is linear in fermionic field goes by the name B-supersymmetrization.
It has received a lot of attention because of their connections with string theories from the point of view of matrix models \cite{Becker}. The B-supersymmetric of the Korteweg-de Vries (KdV) \cite{Becker}, dispersionless two boson hierarchy \cite{Brunellijc}, Sawada-Kotera \cite{Popowi}, modified KdV and Camassa-Holme quations \cite{Choud} have been constructed. In the meanwhile, the methodologies involved in the study of integrable systems have been expanded to the supersymmetric framework \cite{darbl,Carstea, Ghosh, nons, inv}. The soliton solutions of supersymmetric systems have been constructed via many methodologies. However, how to find exact interaction solutions among solitons and other kinds
of complicated waves is an important problem for supersymmetric integrable models.
In this paper, we will use the bosonization \cite{Andrea,lou} and consistent tanh expansion (CTE) \cite{gaox} methods on the ${\cal N} = 1$ supersymmetric mKdV-B (SmKdV-B) system. These interaction solutions among solitons and other types of solitary waves such as Painlev\'{e} II waves and cnoidal waves are explicitly given. All results can be directly transformed to other B-supersymmetrization systems.

The ${\cal N} = 1$ SmKdV-B system reads \cite{Becker}
\begin{equation}\label{smkd}
\Phi_{t} + D^6\Phi - 6 (D \Phi)^2 D^2 \Phi = 0,
\end{equation}
where $D = \partial_\theta + \theta\partial_x$ is the covariant derivative.
The commuting space variable $x$ is extended to a doublet $(x, \theta)$, where $\theta$ is a Grassmann variable.
The Taylor expansion of the superfield with respect to $\theta$ is $\Phi(\theta, x, t)= \xi(x, t) + \theta u(x, t)$.
The component version of \eqref{smkd} thus reads
\begin{subequations}\label{com}
\begin{eqnarray}
& u_{t} - 6 u^2 u_{x} +  u_{xxx} =  0, \\
& \xi_{t} - 6  u^2 \xi_x + \xi_{xxx}= 0.
\end{eqnarray}
\end{subequations}
The bosonic equation (2a) is not depend on the fermionic variable, and equation (2b) is linear in the fermionic field.
To avoid the difficulties in dealing with the anticommutative fermionic field of the supersymmetric equations, the component fields $\xi$ and $u$ are expanded as the following form by introducing additional two fermionic parameters \cite{Andrea,lou,renb,rensuper}
\begin{subequations}\label{ffield}
\begin{eqnarray}
& \xi(x, t)= u_1 \zeta_1 + u_2 \zeta_2, \\
& u(x, t) = u_0 + u_{12} \zeta_1 \zeta_2,
\end{eqnarray}
\end{subequations}
where $\zeta_1$ and $\zeta_2$ are two Grassmann parameters, the coefficients $u_{k}=u_{k}(x,t) \,(k=0, 1, 2)$ and $u_{12}=u_{12}(x,t)$ are four usual real or complex functions with respect to the usual space-time variables. Substituting \eqref{ffield} into \eqref{com} yields the bosonized SmKdV-B (BSmKdV-B) system
\begin{subequations}\label{ffom}
\begin{eqnarray}
& u_{k,t} - 6 u_0^2 u_{k,x} +  u_{k,xxx} = 0, \\
& u_{12,t}- 6 (u_{12} u_{0}^2)_x + u_{12,xxx} = 0.
\end{eqnarray}
\end{subequations}
Equation (4a) is just the usual mKdV equation which has been widely investigated \cite{mkdv}. Equation (4b) is linear homogeneous in $u_{12}$. These pure bosonic systems can be easily solved in principle. This is one of the advantages of the bosonization approach.

The paper is organized as follows. In Section 2, a CTE approach is developed to the BSmKdV-B
equation. It is proved that the BSmKdV-B equation is CTE solvable system. A nonauto-BT theorem is obtained with the CTE method.
In section 3, some novel exact solutions of the BSmKdV-B equation are derived through a nonauto-BT theorem.
The last section is a simple summary and discussion.


\section{CTE method for BSmKdV-B system}

The CTE method is developed to find interaction solutions between solitons and any other types of solitary waves \cite{gaox}. The method has been valid for lots of nonlinear integrable systems \cite{crp,high,Interaction,renin,cwang}.
According to the CTE method \cite{crp,high,Interaction,renin,cwang,reninte}, the expansion solution has the form
\begin{subequations}\label{ff}
\begin{align}
& u_{k} = v = v_{0} + v_{1}\tanh (f), \\
& u_{12} = w = w_{0} + w_{1}\tanh (f) + w_{2}\tanh(f)^2.
\end{align}
\end{subequations}
where $v_{0}$, $v_{1}$, $w_{0}$, $w_{1}$, $w_2$ and $f$ are functions of
$(x, t)$ and should be determined later. Substituting (5a) into (4a) gets
\begin{eqnarray}\label{feq}
& 6f_xv_1( v_1^2 - f_x^2) \tanh(f)^4 + 6(2 f_x v_1^2 v_0+ f_x^2 v_{1,x} + v_1 f_x f_{xx} -v_1^2 v_{1,x}) \tanh(f)^3 - v_1\Bigl(f_t + f_{xxx} - 8f_x^3 + \hspace{0.1cm} \nonumber \\
& 6f_x v_1^2-6 f_x v_0^2 + 6v_1v_{0,x} + 12v_0v_{1,x} + \frac{3(v_{1,x}f_{x})_x}{v_1} \Bigr)\tanh(f)^2 + (v_{1,t}+v_{1,xxx} - 6(v_0^2v_1)_{x} - 6f_xf_{xx}v_1 - \hspace{0.1cm} \nonumber \\
&6f_x^2v_{1,x} - 12f_xv_1^2v_0)\tanh(f) + v_{0,t} + v_{0,xxx} + v_1(f_t+f_{xxx} - 2f_x^3) + 3(v_{1,x}f_x)_x - 6v_0^2(v_{0,x} + v_1f_x)=0. \hspace{0.35cm}
\end{eqnarray}
By setting the coefficients of $\tanh(f)^4$ and $\tanh(f)^3$ in \eqref{feq} to zero, one determines $v_{1}$ and $v_{0}$ as
\begin{align}\label{vlvf}
v_1 = -f_x, \hspace{1cm} v_0= -\frac{f_{xx}}{2f_x}.
\end{align}
By setting the coefficient of $\tanh(f)^2$ in \eqref{feq} to zero and using \eqref{vlvf}, one obtains the equation for $f$
\begin{align}\label{vlf}
f_t = - f_{xxx} + 2 f_x^3 + \frac{3}{2} \frac{f_{xx}^2}{f_x}.
\end{align}
One can verify that the coefficients of $\tanh(f)^1$ and $\tanh(f)^0$ in \eqref{feq} are identically zero by using \eqref{vlf}.
Substituting \eqref{ff} and \eqref{vlvf} into (4b) yields
\begin{eqnarray}\label{geq}
& 12f_x (f_x^2 w_1 + f_x w_{2,x} - f_{xx}w_2) \tanh(f)^4+ 2\bigl( 6f_x^3w_0 - 9 f_xf_{xx}w_1 + (2f_{xxx} - f_t + 8f_x^3 + \frac{3f_{xx}^2}{2f_x})w_2  \nonumber \\
& - 3f_x w_{2,xx} \bigr) \tanh(f)^3 - \bigl( 3f_{xx} (2f_x + \frac{f_{xxx}}{f_x^2} - \frac{f_{xx}^2}{f_x^3})w_2 + (f_t - 5f_{xxx} + 10f_x^3 - \frac{3f_{xx}^2}{2f_x} )w_1 + 18 f_xf_{xx}w_0  \nonumber \\
&  - w_{2,t} - w_{2,xxx} + (24f_x^2 + \frac{3f_{xx}^2}{2f_x^2})w_{2,x} - 3f_{xx}w_{1,x} + 3f_xw_{1,xx} + 6f_x^2w_{0,x} \bigr)\tanh(f)^2 + \bigl( 6(f_{xxx}-2f_x^3)w_0 \nonumber \\
& + 3f_{xx}(2f_x -\frac{f_{xxx}}{f_x^2} + \frac{f_{xx}^2}{f_x^3})w_1 + (2f_t + 2f_{xxx} - 16f_x^3  - \frac{3f_{xx}^2}{f_x})w_2 + w_{1,t} +w_{1,xxx} - (6f_x^2 + \frac{3f_{xx}^2}{2f_x^2}) w_{1,x} \nonumber \\
& + 6(f_{x}w_{2,x})_x - 6f_x^2w_{1,x} \bigr) \tanh(f) + 3f_{xx} (2f_x - \frac{f_{xxx}}{f_x^2} + \frac{f_{xx}^2}{f_x^3}) w_0 + (f_t + f_{xxx} - 2f_x^3 - \frac{3f_{xx}^2}{2f_x}) w_1 \nonumber \\
& + 6f_xf_{xx}w_2 + w_{0,t} + w_{0,xxx} + 3(f_xw_{1,x})_x + 6f_x^2w_{2,x} - \frac{3f_{xx}^2w_{0,x}}{2f_x^2} =0.
\end{eqnarray}
Similarly, by setting the coefficients of $\tanh(f)^4$ and $\tanh(f)^3$ in \eqref{geq} to zero, one determines $w_1$ and $w_0$ as
\begin{align}\label{vlvp}
w_1= \frac{f_{xx}w_2}{f_x^2} -\frac{w_{2,x}}{f_x}, \hspace{0.6cm} w_0 = \Bigl(\frac{f_t}{6f_x^3}-\frac{f_{xxx}}{3f_x^3} + \frac{5f_{xx}^2}{4f_x^4} -\frac{4}{3}\Bigr)w_2 + \frac{w_{2,xx}}{2f_x^2} -\frac{3f_{xx}w_{2,x}}{2f_x^3}.
\end{align}
We denote $w_2$ as $n$ for simplicity. Setting the coefficient of $\tanh(f)^2$ to zero, we reduce the equation for $n$
\begin{eqnarray}\label{gvl}
n_t + n_{xxx} - 6 f_x^2 n_{x} - \frac{6f_{xx} n_{xx} + 3 f_{xxx}n_x}{f_x} + \frac{27f_{xx}^2n_x + 12f_{xx}f_{xxx}n}{2f_x^2} - \frac{12f_{xx}^3n}{f_x^3}=0.
\end{eqnarray}
The coefficients of $\tanh(f)^1$ and $\tanh(f)^0$ in \eqref{geq} are identically zero by using \eqref{gvl}.
While all the coefficients of powers $\tanh(f)$ of the BSmKdV-B system can be vanished by using appropriate $v_{0}$, $v_{1}$, $w_{0}$, $w_{1}$, $w_2$ and $f$,
we call the expansion \eqref{ff} is a CTE and the BSmKdV-B system is CTE solvable \cite{gaox}. In summary, we have the following a nonauto-B\"{a}cklund (BT) theorem for the BSmKdV-B system \eqref{ffom}.

{\bf Nonauto-BT theorem.} The fields $v$ and $w$
\begin{subequations}\label{fvetor}
\begin{align}
v & = - f_x \tanh(f) + \frac{f_{xx}}{2f_x} , \\
w & = n\tanh(f)^2 + \bigl(\frac{f_{xx}n}{f_x^2} - \frac{n_x}{f_x} \bigr)\tanh(f) - \frac{4}{3}n + \frac{n_{xx}}{2f_x^2} - \frac{9f_{xx}n_{x} + 2nf_{xxx} + n f_t}{6f_x^3} + \frac{3f_{xx}^2n}{2f_x^4} ,
\end{align}
\end{subequations}
are a solution of the BSmKdV-B system \eqref{ffom},
while $f$ and $n$ satisfy \eqref{vlf} and \eqref{gvl}.


\section{Interaction solutions of BSmKdV-B system with a nonauto-BT theorem}

Some novel solutions of the BSmKdV-B system can be found by using the above nonauto-BT theorem. Here three examples are listed in the following.

{\bf Example I.} A quite trivial solution of \eqref{vlf} and \eqref{gvl} has the form
\begin{align}\label{solm}
& f = k_1 x+\omega_1 t + l_1, \hspace{1.0cm} n = k_2 x+\omega_2 t + l_2,
\end{align}
where $k_1$, $k_2$, $l_1$ and $l_2$ are arbitrary constants and $\omega_1$, $\omega_2$ are determined by the relations
\begin{align}
\omega_1= 2k_1^3, \hspace{1cm} \omega_2=6k_2k_1^2.
\end{align}
The soliton solution of BSmKdV-B system reads in the following form by using the line solution \eqref{solm} and the nonauto-BT theorem
\begin{subequations}\label{fsolu}
\begin{align}
v & = -k_1 \tanh(k_1 x + 2k_1^3 t+l_1), \\
w & = n\tanh(k_1 x + 2k_1^3t+l_1)^2 -\frac{k_2}{k_1}\tanh(k_1 x + 2k_1^3t) - n.
\end{align}
\end{subequations}
Though the soliton solution \eqref{fsolu} is a traveling wave in the $(x, t)$ for the boson
field $v$, it is not a traveling wave for other boson field $w$ except for the case $n$ being constants, i.e., $k_2=0$.

{\bf Example II.} For the usual mKdV system, there exists a Painlev\'{e} II reduction if one uses
the scaling symmetry \cite{cls}. The scaling group invariant solution $f$ of \eqref{vlf} thus reads
\begin{align}\label{solfs}
f = c\ln(t-b) + W(\xi), \hspace{1.2cm} \xi = \frac{x-a}{(t-b)^{\frac{1}{3}}}.
\end{align}
Substituting \eqref{solfs} into (12a), a second order ordinary differential equation for the field $W_1$ yields
\begin{align}\label{sols}
W_{1,\xi\xi}= 2 W_1^3 + \frac{1}{3} \xi W_1 + \frac{3W_{1,\xi}^{\,2}}{2W_1}  - c, \hspace{1cm} W_1 = W_\xi,
\end{align}
which is the equivalent Painlev\'{e} II reduction \cite{gaox}, $a, b$ and $c$ are constants.
The solution for other field $n$ is obtained by solving
\begin{align}\label{solvv}
& M_{\xi\xi\xi}= \Bigl(\frac{1}{3} \xi + 6W_1^2\Bigr) M_{\xi} + \frac{3W_{1,\xi\xi} M_{\xi} + 6 W_{1,\xi}M_{\xi\xi}}{W_1} \\ \nonumber
&\hspace{0.9cm} - \frac{27W_{1,X}^2M_X + 12W_{1,X} W_{1,XX}M}{2W_1^2} + \frac{12W_{1,X}^{\,3}M}{W_1^3}, \hspace{1.0cm}  n=M(\xi),
\end{align}
The interaction between solitons and Painlev\'{e} II waves of the BSmKdV-B system can be obtained from the nonauto-BT theorem
\begin{subequations}\label{flu}
\begin{align}
v & = \frac{W_1}{(t-b)^\frac{1}{3}} \tanh(f) - \frac{1}{2}\frac{W_{1,\xi}}{(t-b)^\frac{1}{3}W_1}, \\
w & = M\tanh(f)^2 - \frac{M_\xi W_1- MW_{1,\xi}}{W_1^2} \tanh(f) - \frac{4}{3} M + \frac{9M_{\xi\xi} - \xi M}{18W_1^2} \nonumber \\
& \hspace{0.35cm} + \frac{cM-9W_{1,\xi}M_{\xi} - 2MW_{1,\xi\xi}}{6W_1^3} + \frac{5MW_{1,\xi}^{\,2}}{4W_1^4}.
\end{align}
\end{subequations}

{\bf Example III.} To find the interaction solutions between solitons and cnoidal periodic waves of \eqref{vlf},
the solution for the $f$ field assume with one line solution $k_1 x+\omega_1 t$ plus an undetermined traveling
wave $F(k x+\omega t)$
\begin{align}\label{fols}
f = k_1 x+\omega_1 t + F(X),\hspace{1.2cm} X = k x+\omega t.
\end{align}
Substituting \eqref{fols} into \eqref{vlf} yields
\begin{align}\label{fsu}
& F_{1,X}^{\,\,2} - 4F_1^{4} - \bigl(C_1k^3 + \frac{12k_1}{k}\bigr) F_1^{3} - \bigl(3C_1k_1k^2 + \frac{12k_1^2}{k^2} + \frac{ 2\omega}{k^3}\bigr)F_1^{2} \\ \nonumber
& - \bigl(3C_1kk_1^2 + \frac{4k_1^3 + \omega_1}{k^3} + \frac{3k_1\omega}{k^4}\bigr) F_1 - C_1k_1^3 - \frac{k_1(k \omega_1 + k_1\omega)}{k^5} =0, \hspace{0.8cm} F_1=F_X,
\end{align}
with $C_1$ is arbitrary constant. The solution of \eqref{fsu} can be solved out in terms of Jacobi elliptic functions \cite{nonloc}. The solution expressed by \eqref{fols} is thus the explicit interaction solutions between one soliton and cnoidal periodic waves. The field $n$ is obtained by solving the following equation
\begin{align}\label{nsu}
& N_{XXX} - \frac{6k}{kF_1+k_1} F_{1,X} N_{XX} + \Bigl(\frac{27k^2}{2(kF_1 + k_1)^2}F_{1,X}^{\,2} - \frac{3k}{kF_1 + k_1} {F_{1,XX}} - \frac{6(kF_1 + k_1)^2}{k^2} \nonumber \\
& + \frac{\omega}{k^3} \Bigr) N_X +\Bigl(\frac{6k^2}{(kF_1+k_1)^2}F_{1,XX} - \frac{2k}{(kF_1 + k_1)^3}F_{1,X}^{2} \Bigr) F_{1,X} N = 0, \hspace{1.2cm} n=N(X).
\end{align}
As the well known exact Jacobi elliptic functions solutions of \eqref{fsu}, we try to build the mapping and deformation relationship \cite{lou} between the solution for $F_1$ and $N$ by using \eqref{nsu}. In order to get the mapping and deformation relationship, the variable transformation is introduced \cite{lou,renb}
\begin{align}\label{ran}
N(X)=N'(F_{1}(X)).
\end{align}
Substituting the transformation \eqref{ran} into \eqref{nsu} and vanishing $F_{1,X}$ via \eqref{fsu}, equation \eqref{nsu} becomes
\begin{align}\label{coll}
L\bigl(F_1+\frac{k_1}{k}\bigr)^3\frac{d^3N'}{d F_1^3} - \frac{3}{2}M\bigl(F_1+\frac{k_1}{k}\bigr)^2 \frac{d^2 N'}{d F_1^2} + 3M\bigl(F_1+\frac{k_1}{k}\bigr) \frac{d N'}{d F_1} - {3M} N' =0,
\end{align}
where
\begin{align}
L=  4k^4 F_1^4 + k^3(C_1 k^4 + 8k_1) F_1^2  + 2k (C_1k_1k^5+2kk_1^2 + \omega) F_1 + C_1k_1^2 k^5 + k\omega_1+ k_1\omega, \,\, \nonumber \\
M =k^3(C_1k^4-4k_1)F_1^2+2k(C_1k^5k_1-4k k_1^2+2\omega)F_1+C_1k^5k_1^2- 4kk_1^3+3k\omega_1+ k_1 \omega, \nonumber
\end{align}
for simplicity. The solution for \eqref{coll} can be directly obtained by using Maple since \eqref{coll} is the linear ordinary differential equation. The mapping and deformation relation is thus constructed by solving \eqref{coll}
\begin{align}\label{sovl}
N= N' =C_2 k F_1^2 + (C_2k_1 + C_3k)F_1 + C_3k_1,
\end{align}
with $C_2$ and $C_3$ are arbitrary constants.

Substituting \eqref{fols}, \eqref{ran} and \eqref{sovl} into \eqref{fvetor}, the corresponding solution of the BSmKdV-B system reads
\begin{subequations}\label{sovw}
\begin{align}
v & = -(k F_1+k_1) \tanh(f) - \frac{k^2}{2}\frac{F_{1,X}}{k F_1+k_1}, \\
w & = (k F_1+k_1)(C_2F_1 + C_3)\tanh(f)^2 - C_2kF_{1,X} \tanh(f) + \frac{k^2(4C_2kF_1 + 3C_2k_1 +C_3k)}{6(k F_1+k_1)^2}F_{1,XX}  \\
& - \frac{k^3(3C_2kF_1 + 2C_2k_1 +C_3k)}{4(F_1+k_1)^3} F_{1,X}^{\,2} - \frac{(C_2F_1 + C_3)(8k^3F_1^3 + 24 k_1k^2F_1^2 + (24kk_1^2-\omega)F_1+8k_1^3 - \omega_1)}{6(F_1+k_1)^2}. \nonumber
\end{align}
\end{subequations}
To show more clearly of this kind of solution, we give two special cases.

{\bf Case I}. One solution of \eqref{fsu} take as
\begin{align}\label{insf}
F_{1}= a_0 + a_1 \mathrm{sn}(a_2X,m),
\end{align}
where $a_0$, $a_1$ and $a_2$ are constants, $\mathrm{sn}(a_2X,m)$, $\mathrm{dn}(a_2X,m)$ and $\mathrm{cn}(a_2X,m)$ are the usual Jacobi elliptic functions with the modulus $m$. $\hat{S}$, $\hat{D}$ and $\hat{C}$ stand for $\mathrm{sn}(a_2X,m)$, $\mathrm{dn}(a_2X,m)$ and $\mathrm{cn}(a_2X,m)$ respectively for simplicity. Substituting \eqref{insf} into \eqref{fsu} and vanishing all the coefficients of different powers of $\hat{S}$, a nontrivial solution is obtained for constants
\begin{align}
& a_0 = - \frac{C_1 k^3}{4} - \frac{3a_2}{2}, \hspace{1.1cm} a_1 = -\frac{a_2m}{2}, \hspace{1.1cm} k_1 = \frac{k(C_1k^3+8a_2)}{4}, \nonumber \\
& \omega= \frac{ k^3 a_2^2 (m^2-5) }{2},  \hspace{1.2cm} \omega_1 = \frac{k^3 a_2^2 (32a_2 + 5C_1 k^3 - C_1k^3 m^2) }{8}.
\end{align}
In this case, the interaction solution of of the BSmKdV-B system becomes by using \eqref{sovw}
\begin{subequations}\label{finf}
\begin{align}
& v = -\frac{1}{2}ka_2(1+m \hat{S}) \tanh(f) - \frac{a_2km}{2(1 + m\hat{S})} \hat{C} \hat{D}, \\
& w = \frac{1}{8} a_2 k (1+m \hat{S})(2C_2a_2m \hat{S} + 4C_3 -6C_2a_2 - C_1C_2k^3)\tanh(f)^2 + \frac{1}{2}C_2a_2^2km \hat{C} \hat{D} \tanh(f) \nonumber \\
& \hspace{0.35cm} + \frac{1}{8}a_2k (C_1C_2k^3-4C_3)(1+m \hat{S}) + \frac{3}{4}a_2^2C_2km \hat{S} + \frac{a_2^2C_2k(m^3\hat{S}^3 - 2m^2 + 3)}{4(1+m \hat{S})},\\
&  \hspace{1.35cm} \Bigl(f=\frac{ka_2}{2}x+ \frac{a_2^3k^3(3m^2+1)t}{4}t - \frac{1}{2} \ln( \hat{D} - m \hat{C})\Bigr). \nonumber
\end{align}
\end{subequations}
The figure 1 plots the interaction solutions between solitons and cnoidal waves for the field $v$ and $w$ respectively. The parameters are $k=1,\, m=0.3,\, a_2=1, k_1=1, C_1=1, C_2=1, C_3=1$.
\input epsf
\begin{figure}
\centering
\rotatebox{270}{\includegraphics[width=5.15cm]{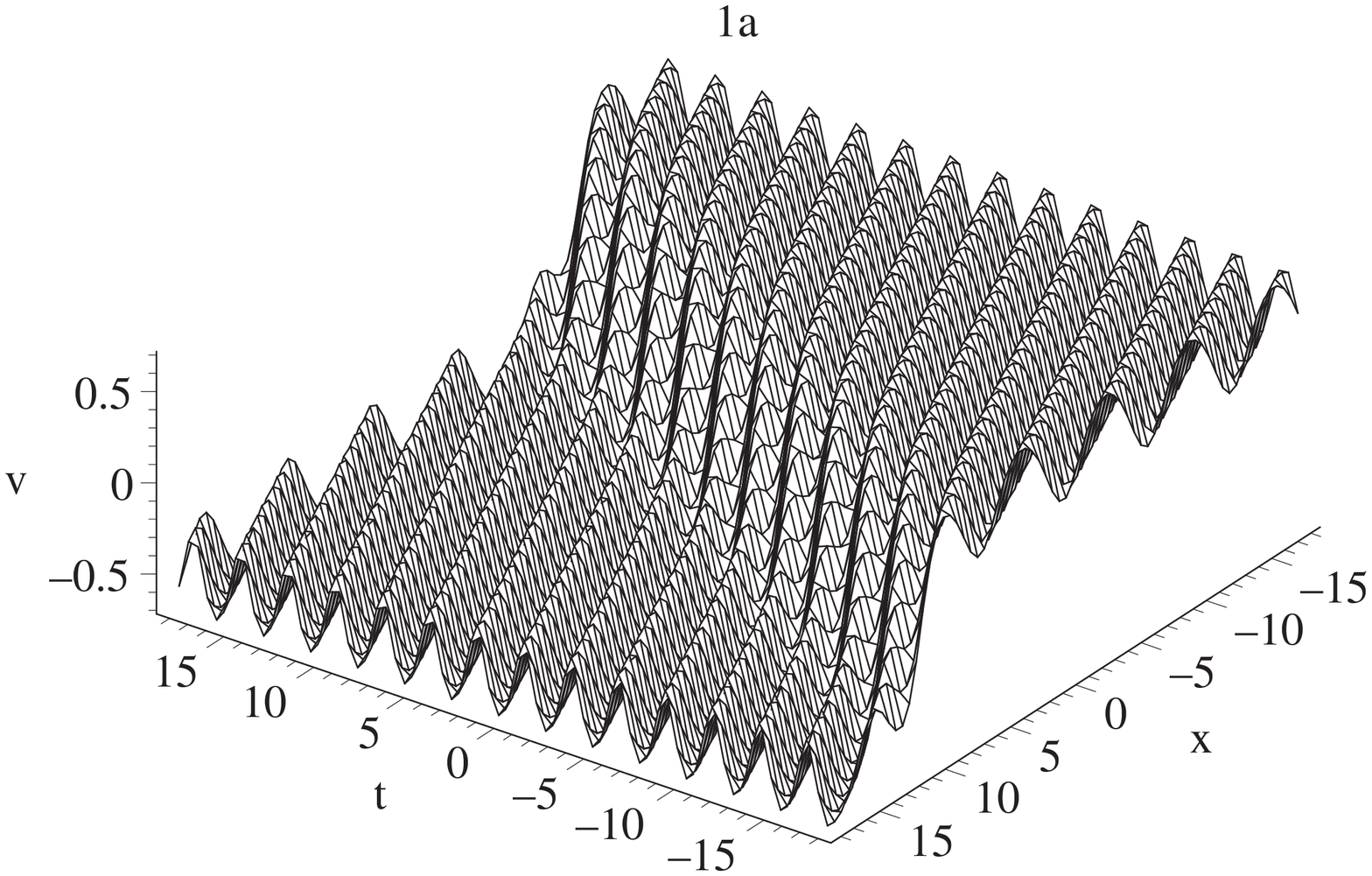}}
\rotatebox{270}{\includegraphics[width=5.15cm]{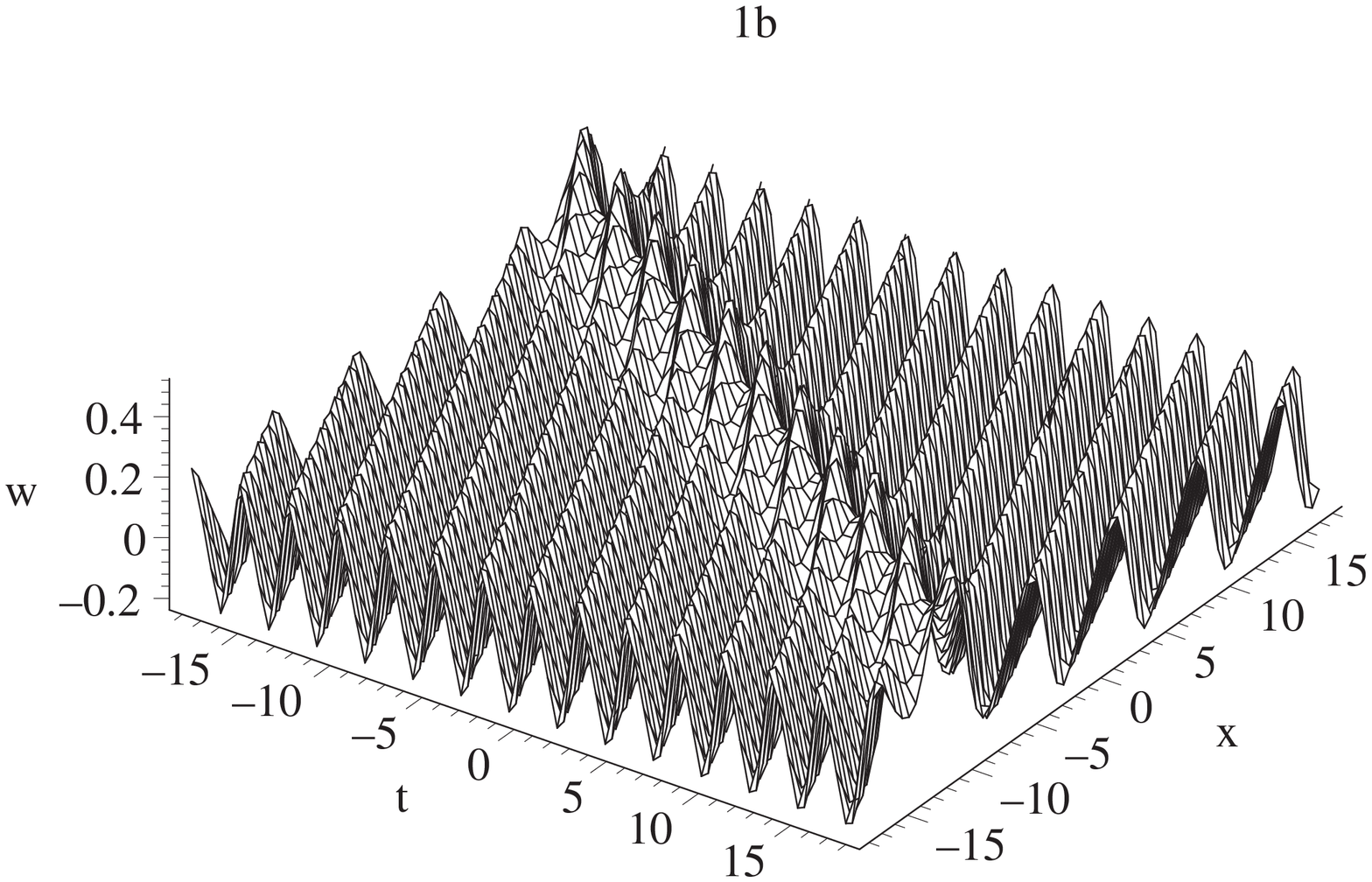}}
\caption{(a) Plot of one kink soliton on the cnoidal wave background expressed by (29a). (b) Plot of the first special soliton-cnoidal wave interaction solution by (29b). The parameters are $k=1,\, m=0.3,\, a_2=1, C_1=1, C_2=1, C_3=1$.}
\end{figure}

{\bf Case II.} Another special solution of \eqref{fsu} reads
\begin{align}\label{inteff}
F_{1}= - b_1b_2 + \frac{b_1b_2}{1 - n\mathrm{sn}(b_1 X,m)^2},
\end{align}
where $b_1, b_2, n$ and $m$ are constants, $\mathrm{sn}(b_1X,m)$, $\mathrm{dn}(b_1X,m)$ and $\mathrm{cn}(b_1X,m)$ are the usual Jacobi elliptic functions. Here, $\bar{S} $, $\bar{D}$ and $\bar{C}$ stand for $\mathrm{sn}(b_1X,m)$, $\mathrm{dn}(b_1X,m)$ and $\mathrm{cn}(b_1X,m)$ respectively for similarity. Substituting \eqref{inteff} into \eqref{fsu} and vanishing all the coefficients of different powers of $\bar{S}$ yields constraints of constants
\begin{align}
& b_1 = \frac{C_1k^3 \sqrt{n(1-n)(m^2-n)}}{4(m^2n-2m^2+n)}, \hspace{0.45cm} b_2 = \sqrt{\frac{(1-n)(m^2-n)}{n}}, \hspace{0.45cm} k_1 = \frac{C_1k^4(1-n)(m^2-n)}{4(m^2n-2m^2+n)},  \\
& \omega = \frac{C_1^2k^9 (1-n)(n-m^2) (m^2n-3m^2+n)}{8(m^2n-2m^2+n)^2}, \hspace{0.35cm} \omega_1 = \frac{C_1^3k^{12} (m^2 n - m^2 - n^2 +n)^2 (m^2-m^2n-n)}{32(m^2n-2m^2+n)^3}.\nonumber
\end{align}
In this case, a interaction solution for the BSmKdV-B system is obtained as
\begin{subequations}\label{finff}
\begin{align}
& v = \frac{C_1k^4(n-1)(m^2-n)}{4(n\bar{S}^2-1) (m^2n-2m^2+n)} \tanh(f) - \frac{nC_1k^4\sqrt{(m^2-n)(1-n)}}{4(m^2n-2m^2+n)(n\bar{S}^2-1)}\bar{S} \bar{C} \bar{D},  \\
& w = \Bigl(\frac{C_1C_3k^4(n-1)(m^2-n)}{4(n\bar{S}^2 -1)(m^2n-2m^2+n)}+\frac{nC_1^2C_2k^7(n-1)^2(m^2-n)^2}{16(n\bar{S}^2 -1)^2(m^2n-2m^2+n)^2} \bar{S}^2\Bigr) \tanh(f)^2\\ &  \nonumber \hspace{0.5cm} +\frac{C_1^2C_2k^7n(m^2-n)^\frac{3}{2}(1-n)^\frac{3}{2}}{8(n\bar{S}^2-1)^2(m^2n-2m^2+n)^2}\bar{S} \bar{C} \bar{D} \tanh(f) + \frac{C_1C_3k^4(1-n)(m^2-n)}{4(n\bar{S}^2 -1)(m^2n-2m^2+n)} \\ \nonumber
&  \hspace{0.5cm} +\frac{C_1^2C_2 k^7n(n-1)(m^2-n)(m^2n^2\bar{S}^6 - 3\bar{S}^4m^2n + \bar{S}^2m^2n +\bar{S}^2m^2+\bar{S}^2m^2+\bar{S}^2n-n)}{16(n\bar{S}^2 -1)^2(m^2n-2m^2+n)^2},\\
&  \hspace{0.6cm} \Bigl(f= \frac{C_1^3m^2k^{12}(1-n)^2(m^2-n)^2t}{16(m^2n-2m^2+n)^3} -\sqrt{\frac{(1-n)(m^2-n)}{n}}E_{\pi}(\bar{S},n,m)\Bigr), \nonumber
\end{align}
\end{subequations}
where $E_{\pi}$ is the third type of incomplete elliptic integral.
The figure 2 plots the interaction solutions between solitons and cnoidal waves expressed by \eqref{finff} with the parameters $k=1,\, m=0.8,\, n=0.2,\, C_1=3, C_2=1, C_3=1$. There are some typical nonlinear waves such as interaction solutions between solitary waves and cnoidal periodic waves in the ocean \cite{nonloc,appl,pplc}. Solutions \eqref{finf} and \eqref{finff} may be useful for studying these types of ocean waves \cite{appl,pplc}.

{\bf Remark.} For a given solution $v$ and $w$ of BSmKdV-B system, the interaction solutions for component fields $\xi$ and $u$ of SmKdV-B can be constructed by using \eqref{ffield}.
The bosonization approach can thus effectively avoid difficulties caused by anticommutative fermionic fields for supersymmetric nonlinear systems.
\input epsf
\begin{figure}
\centering
\rotatebox{270}{\includegraphics[width=5.15cm]{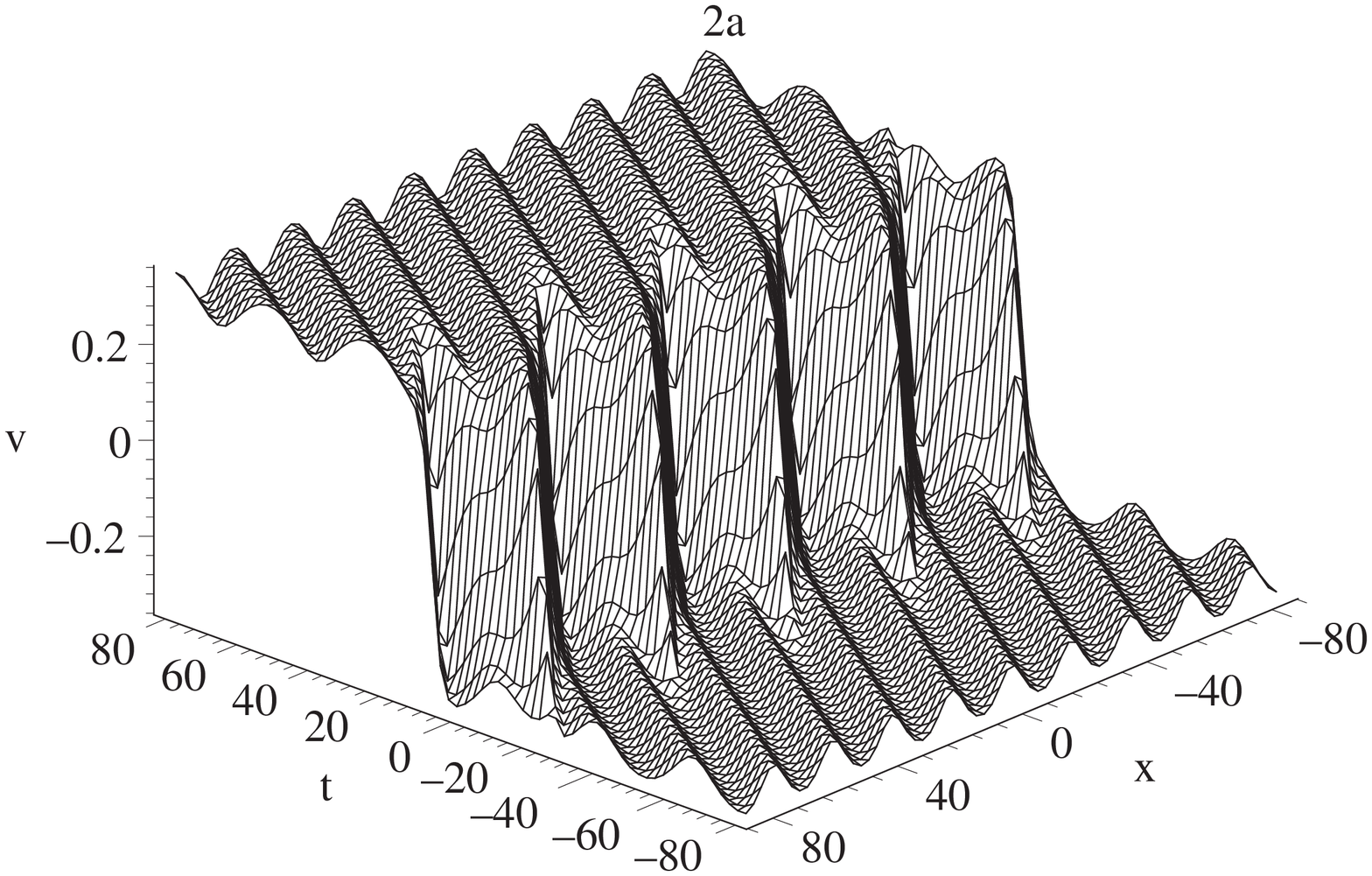}}
\rotatebox{270}{\includegraphics[width=5.15cm]{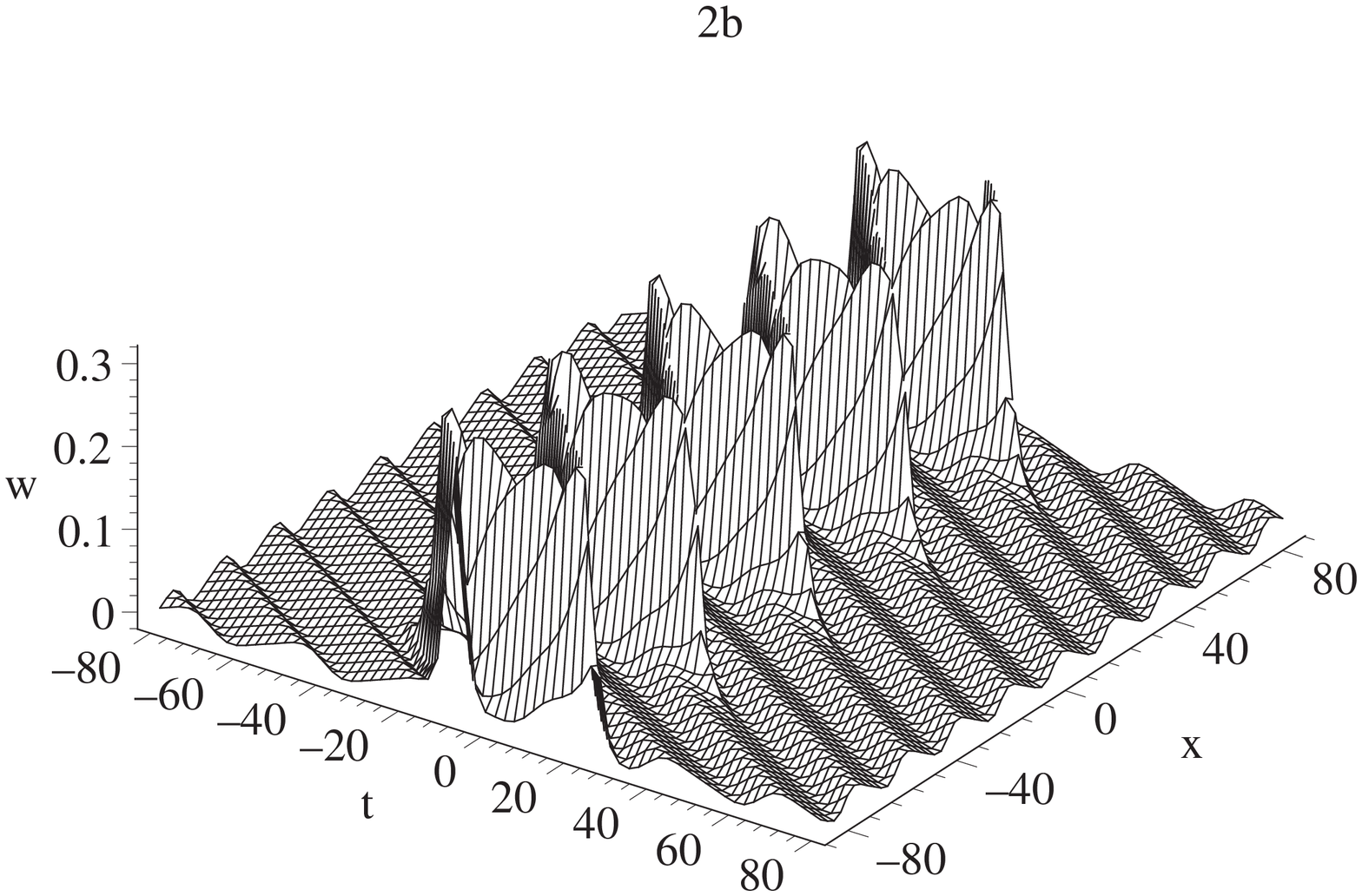}}
\caption{(a) Plot of one kink soliton on the cnoidal wave background expressed by (31a).
(b) Plot of a second special soliton-cnoidal wave interaction solution by (31b). The parameters are $k=1,\, m=0.8,\, n=0.2,\, C_1=3, C_2=1, C_3=1$.}
\end{figure}

\section{Conclusions}

In summary, the ${\cal N} =1$ SmKdV-B system is mapped to a system of coupled bosonic equations by means of the bosonization approach.
The BSmKdV-B system is just the usual mKdV equation together with a linear differential equation.
Then, the CTE approach is applied to the BSmKdV-B equation. It is proved that the BSmKdV-B equation is CTE solvable system. A nonauto-BT theorem is obtained with the CTE method. Various explicit solutions of the BSmKdV-B system such as soliton-Painlev\'{e} II waves and soliton-cnoidal waves are obtained by using the nonauto-BT theorem. For the interaction soliton-cnoidal waves, two cases are given both in analytical and graphical ways via combining the mapping and deformation method. This kind of interaction solutions may be useful in real physical phenomena.

For the bosonization approach, we can also introduce $N\geq 2$ fermionic parameters $\zeta_i \,(i=2,3,...,N)$ to study the B-supersymmetrization of nonlinear evolution system. On the other hand, lots of B-supersymmetrization systems are introduced from the usual action principle \cite{Choud}. The interaction solutions among solitons and other complicated waves for these B-supersymmetrization systems are worth studying.

{\bf Acknowledgment}:

\noindent
This work was partially supported by the National Natural Science Foundation of China under Grant Nos. 11305106, 11275129 and 11347183.


\end{document}